\documentclass[twocolumn,aps,floatfix]{revtex4}
\usepackage{dcolumn,graphicx}
\def\be {\begin{equation}}
\def\ee {\end{equation}}
\def\mn {{\mu\nu}}
\def\ba {\begin{eqnarray}}
\def\ea {\end{eqnarray}}
\def\cm {{\cal M}}
\def\cl {{\cal L}}

\def\del {\partial}

\def\om {\omega}

\def\vq {\vec q}
\def\vk {\vec k}
 
\def\de {\delta}                                 
\def\Gm {\Gamma}
\def\sg {\sigma}
\def\omp {\om_\pi}
\def\omh {\om_h}
\def\nn {\nonumber}

\begin{document}
\title{Medium effects on the thermal conductivity of a hot pion gas}
\author{Sukanya Mitra}
\author{Sourav Sarkar}
\affiliation{Theoretical Physics Division, Variable Energy Cyclotron Centre,
 1/AF Bidhannagar Kolkata - 700064, India}

\begin{abstract}
We investigate the effect of the medium on the thermal conductivity of a pion 
gas out of chemical equilibrium by solving the relativistic transport equation
in the Chapman-Enskog and relaxation time approximations. Using an effective 
model for the $\pi\pi$ cross-section involving $\rho$ and $\sigma$ meson 
exchange, medium effects are incorporated through thermal one-loop 
self-energies. The temperature dependence of the thermal conductivity is 
observed to be significantly affected.
\end{abstract}

\maketitle

The observation of large elliptic flow of hadrons in heavy ion collisions at
RHIC has led to the description of quark-gluon plasma as a nearly perfect 
fluid~\cite{Csernai}.
This interpretation is based on the small but finite value of the shear
viscosity to entropy density ratio required in a relativistic hydrodynamic
description of the collision. The effects of dissipation on the dynamical
evolution of matter produced in relativistic heavy ion collisions have thus
been a major topic of discussion in recent times~\cite{QMProc}.
At the microscopic level dissipative phenomena are studied by considering
small departures from equilibrium.
In kinetic theory the transport of momenta and heat as a result of
collisions is quantitatively expressed in terms of coefficients 
of viscosity and thermal conductivity~\cite{deGroot,Zubarev}. A large number of
studies on the viscous coefficients have been performed in the transport
approach. The shear viscosity
$\eta$ has been most commonly discussed followed by the bulk viscosity $\zeta$,
both for partonic as well as hadronic systems~\cite{Gavin,Prakash,Davesne,
Santalla,Dobado3,Chen,Itakura,Kharzeev,Fraile1,Dobado1,Chen2,Noronha,
Demir,Redlich,Fraile3,Dobado2,Cassing,Rincon,Fraile2,Greif,Denicol}. 
The interesting issue concerning the behaviour of the viscosities in the
vicinity of the transition from partonic to hadronic matter have also been
discussed~\cite{Csernai,Redlich,Kharzeev,Dobado1,Chen2,Fraile3,Dobado2}. 
While the value of $\eta/s$ is expected to go through a minimum near the critical
temperature~\cite{Csernai,Redlich}, $\zeta/s$ is believed to be large or
diverging~\cite{Kharzeev,Chen2,Fraile3} at or near the transition.

The effects of heat flow in heavy ion collisions has received much less
attention. This is presumably on account of the fact that the net baryon number
in the central rapidity region at the RHIC and LHC is very small. However, 
at FAIR energies or in the low energy runs at RHIC
the baryon chemical potential is expected to be significant and 
heat conduction by baryons may play a more important role. 
On the other hand, a thermal system consisting of pions can sustain heat 
conduction despite the fact that the pions themselves do not carry 
baryon number~\cite{Gavin}. This is due to the fact that the total number of 
pions in heavy ion collisions is essentially conserved.
Pion number changing reactions are not sustained towards the late stages 
where collisions are mostly elastic and the system undergoes 
chemical freezeout. As the system expands and cools a pion chemical potential 
develops in order to keep the pion number fixed. Based on such a scenario
a few studies of heat conduction by pions have been carried out. Using the
experimental $\pi\pi$ cross-section the thermal conductivity of a pion gas was
estimated in~\cite{Gavin,Prakash,Davesne} whereas in~\cite{Rincon} a
unitarized scattering amplitude was employed. The heat conductivity was also
obtained using the Kubo formula in~\cite{Fraile1,Fraile2,SS_AHEP}. For the case
of a classical gas, heat flow has been studied recently in a transport 
model~\cite{Greif} and a fluid-dynamical theory was derived~\cite{Denicol}.
Investigating the effect of thermal conductivity on first order phase 
transitions, non-trivial fluctuation effects were observed in~\cite{Skokov}
which may result in a non-monotonic behaviour of certain observables as a
function of collisional energy and may be seen from experimental analysis at
RHIC and FAIR. A clear picture of the behaviour of thermal conductivity in
the vicinity of a phase transition is however yet to emerge.

In the kinetic theory approach the dynamics of interaction resides in 
the differential cross-section which goes as an input. In
almost all estimations of the transport coefficients a
vacuum cross-section was employed. In~\cite{Sukanya1,Sukanya2} a
medium dependent cross-section was used in the evaluation of shear and bulk viscosities
of a pion gas which resulted in a significant deviation from the results 
obtained with the $\pi\pi$ cross-section in vacuum. 

In this work we study the 
temperature dependence of the thermal conductivity of a pion gas. 
In particular, our intention is to emphasize on the effect of the medium on its
temperature dependence brought in by the cross-section. To this end we employ
an effective Lagrangian approach in which
the $\pi\pi$ scattering amplitude is obtained in terms of $\rho$ and 
$\sigma$ meson exchange.
Medium effects are then incorporated by introducing
in-medium propagators dressed by one loop self energies calculated
in the framework of thermal field theory. We use a temperature dependent pion
chemical potential and obtain the thermal conductivity for
temperatures in the range between chemical and kinetic freezeout in heavy ion
collisions.

The thermal conductivity $\lambda$ is obtained by solving the Uehling-Uhlenbeck 
equation in the Chapman-Enskog approximation to first order. This 
calculation is performed along the lines of~\cite{Polak,Davesne} and is
described elaborately in~\cite{Sukanya2}. Here we provide only the basics 
of the formalism. We start with the transport equation for the phase-space 
distribution $f(x,p)$ of a relativistic pion gas which is given by
\be
p^\mu\partial_\mu f(x,p)=C[f]~.
\label{treq}
\ee
For binary elastic collisions $p+k\to p'+k'$, the collision term $C[f]$ is
defined by,
\ba
C[f]=&&\frac{1}{2}\int d\Gamma_k\ d\Gamma_{p'}\ d\Gamma_{k'}[f(x,p')f(x,k') \{1+f(x,p)\}\nonumber\\
&&\times\{1+f(x,k)\}-f(x,p)f(x,k)\{1+f(x,p')\}\nonumber\\
&&\times\{1+f(x,k')\}]\ W
\ea
where, 
\[ 
W=(2\pi)^4\delta^4(p+k-p'-k')\frac{1}{2}|\cm|^2,\,\,\,
d\Gamma_q=\frac{d^3q}{(2\pi)^3 2E_q}~.
\]

For a pion gas slightly away from equilibrium 
the phase space distribution function can be expanded in the first Chapman-Enskog
approximation as
\be
f(x,p)=f^{(0)}(x,p)+f^{(0)}(x,p)[1+f^{(0)}(x,p)]\phi(x,p),
\label{ff}
\ee
where, $f^{(0)}(x,p)=[e^{\frac{p\cdot u(x)-\mu(x)}{T(x)}}-1]^{-1}$ is the 
local equilibrium Bose distribution function. The deviation function $\phi(x,p)$
then satisfies the following linearized transport equation
\be
p^\mu\partial_\mu f^{(0)}(x,p)=-\cl[\phi]~
\label{treq2}
\ee
in which the collision term is given by,
\ba
\cl[\phi]&=&f^{(0)}(x,p)\frac{1}{2}\int d\Gamma_k\ d\Gamma_{p'}\ d\Gamma_{k'}f^{(0)}(x,k)\nonumber\\
&\times&\{1+f^{(0)}(x,p')\}\{1+f^{(0)}(x,k')\}\nonumber\\
&\times&[\phi(x,p)+\phi(x,k)-\phi(x,p')-\phi(x,k')]\ W~.
\ea 
To solve this equation $\phi$ is generally
expressed in the form
\be
\phi=A\partial_{\nu}u^{\nu}+B_{\mu}\Delta^{\mu\nu}(T^{-1}\partial_{\nu}T
-Du_{\nu})-C_{\mu\nu}\langle \partial^{\mu} u^{\nu} \rangle
\label{phi}
\ee
where $D=u^\mu\del_\mu$ and $\Delta^{\mu\nu}=g^{\mu\nu}-u^\mu u^\nu$, $u^\mu$ being
the flow velocity.
The scalar and tensor processes denoted by the first and third terms are
connected with bulk and shear viscosities respectively. The vector process 
given by the second term corresponds to the transport phenomena related to 
thermal conduction.
Comparing with the expression for energy 4-flow, 
$I^{\mu}=\lambda(\partial_\sigma T-TDU_\sigma)\Delta^{\mu\sigma}$
the coefficient of thermal conductivity $\lambda$ can be defined as
\be
\lambda=\frac{2}{3T}\int d\Gamma_p f^{(0)}(1+f^{(0)})B_{\nu}p^\nu  (p\cdot u -h)
\ee
where $h$ is the enthalpy per particle. 
The unknown coefficient $B_\mu=B\Delta_ {\mu\nu} p^\nu$ can be obtained by
solving the equation,
\be
\cl[B_\mu]=-\frac{1}{T}f^{(0)}(1+f^{(0)})\Delta_{\mu\nu} p^\nu(p\cdot u -h)~.
\ee 
Here we follow the procedure outlined in~\cite{Polak,Davesne} in which
$B_\mu$ is expanded in terms of orthogonal Laguerre
polynomials of order $3/2$. After some simplifications
(discussed in detail in Refs.~\cite{Davesne,Sukanya2})
the first approximation to thermal conductivity comes out to be,
\be
\lambda=\frac{T}{3m_\pi}\frac{\beta_1^2}{b_{11}}
\ee
where,
\ba
\beta_1&=&-3z^2\left\{1+5z^{-1} \frac{S^{2}_3(z)}{S^{1}_2(z)}-\left(\frac{S^{1}_3(z)}
 {S^{1}_2(z)}\right)^2\right\}\nonumber\\
b_{11}&=&I_1(z)+I_2(z)~~{\rm with}~~z=m_\pi/T~.
\label{beta-b11}
\ea
The integrals $I_\alpha(z)$ are given by~\cite{Sukanya1,Sukanya2}
\ba
I_\alpha(z)&=&\frac{8z^5}{[S_2^{1}(z)]^2} \ e^{(-2\mu_\pi/T)}\int_0^\infty d\psi\ \cosh^3\psi\sinh^7\psi\nonumber\\
&\times&\int_0^\pi d\Theta\sin\Theta\{\frac{1}{2}\frac{d\sigma}{d\Omega}\}(\psi,\Theta)\int_0^{2\pi}d\phi\nonumber\\
&\times&\int_0^\infty d\chi \sinh^{(2\alpha +2)} \chi\int_0^\pi d\theta\sin\theta\nonumber\\
&\times&\frac{e^{2z\cosh\psi\cosh\chi}}{(e^E-1)(e^F-1)(e^G-1)(e^H-1)}\ M_\alpha(\theta,\Theta),\nonumber\\
&&
\label{c00_bose}
\ea
and $S_n^\alpha(z)$ denotes integrals over Bose functions which can be
expressed in terms of infinite series as
$S_n^\alpha(z)=\displaystyle\sum_{k=1}^\infty e^{k\mu/T} k^{-\alpha} K_n(kz)$, 
$K_n(x)$ denoting the modified Bessel function of order $n$.
The exponents in the Bose functions and the functions $M_\alpha(\theta,\Theta)$
are respectively given by
\ba
E&=&z(\cosh\psi\cosh\chi-\sinh\psi\sinh\chi\cos\theta)-\mu_\pi/T\nonumber\\
F&=&z(\cosh\psi\cosh\chi-\sinh\psi\sinh\chi\cos\theta')-\mu_\pi/T\nonumber\\
G&=&E+2z\sinh\psi\sinh\chi\cos\theta\nonumber\\
H&=&F+2z\sinh\psi\sinh\chi\cos\theta'~,
\ea
\ba
M_1(\theta,\Theta)&=&\cos^2\theta+\cos^2\theta'
-2\cos\theta\cos\theta'\cos\Theta~,\nonumber\\
M_2(\theta,\Theta)&=&[\cos^2\theta-\cos^2\theta']^2
\ea
where
\(
\cos\theta'=\cos\theta\cos\Theta-\sin\theta\sin\Theta\cos\phi~.
\)

\begin{figure}
\includegraphics[scale=0.4]{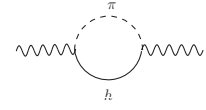}
\caption{The self-energy diagrams for
$h=\pi,\om,h_1,a_1$ mesons.}
\label{one_loop}
\end{figure}

The $\pi\pi$ cross-section is the key dynamical input 
for evaluating transport coefficients. Here the scattering is assumed 
to proceed via $\sigma$ and $\rho$ meson exchange in the medium.
From the effective interaction~\cite{Vol_16} 
\be
\cl=g_\rho\vec\rho^\mu\cdot\vec \pi\times\del_\mu\vec \pi+\frac{1}{2}g_\sigma
m_\sigma\vec \pi\cdot\vec\pi\sigma
\ee
the matrix elements for $\pi\pi$ scattering
 are given by the following expressions where 
the widths of the $\sigma$ and $\rho$ mesons have been introduced
in the propagators involved in the corresponding
$s$-channel processes. We thus have
\ba
\cm_{I=0}&=&2g_\rho^2\left[\frac{s-u}{t-m_\rho^2}+\frac{s-t}{u-m_\rho^2}\right]\nonumber\\
&+&g_\sigma^2 m_\sigma^2\left[\frac{3}{s-m_\sg^2+im_\sg\Gm_\sg}+\frac{1}{t-m_\sg^2}+
\frac{1}{u-m_\sg^2}\right]\nonumber\\
\cm_{I=1}&=&g_\rho^2\left[\frac{2(t-u)}{s-m_\rho^2+im_\rho\Gamma_\rho}+
\frac{t-s}{u-m_\rho^2}-\frac{u-s}{t-m_\rho^2}\right]\nonumber\\   
&+&g_\sigma^2 m_\sigma^2\left[\frac{1}{t-m_\sg^2}-\frac{1}{u-m_\sg^2}\right]~.
\label{amp}
\ea
Defining the isospin averaged amplitude as 
$|\cm|^2=\frac{1}{9}\sum_I|\cm_I|^2$ and ignoring
the non-resonant $I=2$ contribution, the cross-section is found to
agree very well~\cite{Sukanya1,Sukanya2} with the estimate based
on measured phase-shifts given in~\cite{Prakash}. In this way it is ensured
that the dynamical model is normalized against experimental data although, 
this approach of introducing the width is
not quite in agreement with low energy theorems based on chiral symmetry. 

To obtain the in-medium cross-section we replace the vacuum width in the 
above expressions by the ones
in the medium. The width is related to the imaginary part of the self-energy 
through the relation~\cite{Bellac}
\be
\Gamma(T,M)=-M {\rm Im} \Pi(T,M)
\ee
where $\Pi$ denotes the one-loop self energy diagrams shown in 
fig.~\ref{one_loop}
and are evaluated using the 
real-time formalism of thermal 
field theory. The $\sigma$ meson self-energy is obtained from 
the $\pi\pi$ loop diagram
whereas in case of the $\rho$ meson the
$\pi\pi$, $\pi\omega$, $\pi h_1$, $\pi a_1$ graphs
are evaluated using interactions from chiral perturbation
theory~\cite{Ecker}. 
The longitudinal and transverse parts of the $\rho$ self-energy
are defined as~\cite{Ghosh}
\be
\Pi^T=-\frac{1}{2}(\Pi_\mu^\mu +\frac{q^2}{\bar q^2}\Pi_{00}),~~~~
\Pi^L=\frac{1}{\bar q^2}\Pi_{00} , ~~~\Pi_{00}\equiv u^\mu u^\nu \Pi_{\mn}~.
\label{pitpil}
\ee
The momentum dependence
being weak we take an average over the 
polarizations.
The imaginary part of the self-energy
obtained by evaluating the loop diagrams is given by~\cite{Mallik_RT}
\ba
&&{\rm Im} \Pi(q_0,\vq)=-\pi\int\frac{d^3 k}{(2\pi)^3 4\om_\pi\om_h}\times\nonumber\\
&& \left[N_1\{(1-f^{(0)}(\omp)-f^{(0)}(\omh))\de(q_0-\om_\pi-\om_h)\right.\nonumber\\
&&+(f^{(0)}(\omp)-f^{(0)}(\omh))\de(q_0-\om_\pi+\om_h)\}+\nonumber\\
&&  N_2\{(f^{(0)}(\omh)-f^{(0)}(\omp))\de(q_0+\om_\pi-\om_h)\nonumber\\
&&\left.-(1-f^{(0)}(\omp)-f^{(0)}(\omh))\de(q_0+\om_\pi+\om_h)\}\right]
\label{ImPi_a}
\ea
where $f^{(0)}(\om)=\frac{1}{e^{(\om-\mu_\pi)/T}-1}$ is the Bose distribution
function with arguments $\om_\pi=\sqrt{\vk^2+m_\pi^2}$ and
$\om_h=\sqrt{(\vq-\vk)^2+m_h^2}$. The terms $N_1$ and $N_2$ stem from the
vertex factors and the numerators of vector propagators, details of which 
can be found in~\cite{Mallik_RT}. The angular integration is
done using the $\de$-functions which 
define the kinematic domains for occurrence of scattering and decay processes
which lead to loss or gain of $\rho$ (or $\sigma$) mesons in the medium. To
account for the substantial $3\pi$ and $\rho\pi$ branching ratios of the heavy
particles in the loop the self-energy function is convoluted with their widths,
\ba
\Pi(q,m_h)&=& \frac{1}{N_h}\int^{(m_h+2\Gm_h)^2}_{(m_h-2\Gm_h)^2}dM^2\,\times\nn\\
&&\frac{1}{\pi} {\rm Im} \left[\frac{1}{M^2-m_h^2 + iM\Gm_h(M) } \right] \Pi(q,M) 
\ea
with
\ba 
N_h&=&\displaystyle\int^{(m_h+2\Gm_h)^2}_{(m_h-2\Gm_h)^2}dM^2\,\times\nn\\
&&\frac{1}{\pi} {\rm Im}\left[\frac{1}{M^2-m_h^2 + iM\Gm_h(M)} \right]~.
\ea
The contribution from the loops with these unstable particles can thus be looked upon as
multi-pion effects in $\pi\pi$ scattering. 
\begin{figure}
\includegraphics[scale=0.28]{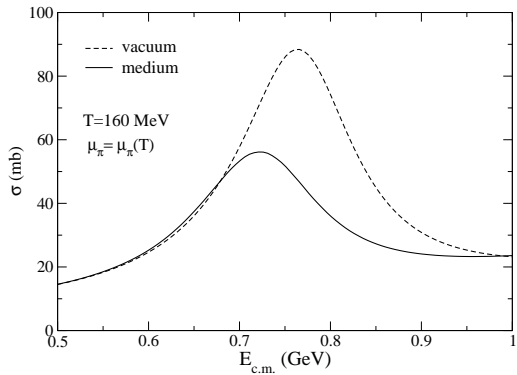}
\caption{The $\pi\pi$ cross-section as a function of centre of mass energy. The
dashed and solid lines respectively indicate the cross-section obtained using the vacuum 
and in-medium widths of the
$\rho$ and $\sigma$ mesons.}
\label{sigmafig}
\end{figure}

It is generally accepted~\cite{Bebie} that the hadronic gas produced after the
transition is in chemical equilibrium where the chemical
potential of pions for example is zero. Chemical freezeout for an evolving 
hadronic gas occurs much earlier than kinetic freezeout. The number-changing 
inelastic collisions cease at chemical freezeout and
the total pion number becomes fixed. Thereafter only elastic collisions take 
place until the pions actually decouple later at kinetic freezeout. The 
pion chemical potential consequently grows from zero to a maximum at 
kinetic freezeout so as to keep the total number of pions fixed.
Here the temperature-dependent pion
chemical potential is taken from Ref.~\cite{Hirano} which implements the above
scenario and is parametrized as
\be
\mu_\pi(T)=a+bT+cT^2+dT^3
\ee
with $a=0.824$, $b=3.04$, $c=-0.028$, $d=6.05\times 10^{-5}$ and $T$, $\mu_\pi$ in MeV. 

We now plot in fig.~\ref{sigmafig}
the total $\pi\pi$ cross-section defined by $\sigma(s)=\frac{1}{2}\int 
d\Omega\frac{d\sigma}{d\Omega}$ with
$\frac{d\sigma}{d\Omega}=\frac{|\cm|^2}{64\pi^2 s}$. The increase in the widths
of the exchanged $\rho$ and $\sigma$ on account of thermal emission and
absorption is reflected in a significant change in both the magnitude and shape of the
cross-section as a function of the c.m. energy. A rough estimate of the mean
free path of pions using the peak value of the in-medium cross-section comes out
to be $\sim$ 1-2 fm at $T=160$ MeV. A macroscopic length scale such as the
typical size of the system at this stage being much larger justifies the use of the 
Chapman-Enskog method for solving the transport equation.

\begin{figure}
\includegraphics[scale=0.4]{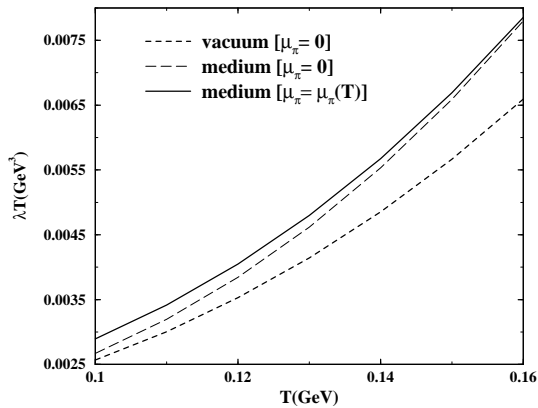}
\caption{$\lambda T$ as a function of $T$ for $\pi\pi$ cross-section
in vacuum and in medium evaluated in the Chapman-Enskog approximation.}
\label{muT}
\end{figure}

We next turn to the results of thermal conductivity. In fig.~\ref{muT} we plot $\lambda T$ as a
function of $T$ evaluated in the Chapman-Enskog approach. The dashed line 
shows results where the vacuum cross-section is used in the integrals
(\ref{c00_bose}). 
For a vanishing pion chemical potential this result 
agrees with those of~\cite{Prakash,Davesne}. Replacing the vacuum widths
by the in-medium widths in the $\rho$ and $\sigma$ propagators in the scattering
amplitudes results in the long dashed line. A substantial medium effect is seen
even for $\mu_\pi=0$ and this is seen to increase with increase of temperature.
We now introduce the temperature dependent $\mu_\pi$ both in the cross-section
and elsewhere in eqs.~(\ref{beta-b11}) and (\ref{c00_bose}). 
This yields the solid line. On comparing with
the long-dashed line the effect of chemical freeze-out is seen to be more 
at lower temperatures since the value of $\mu_\pi(T)$ increases as one
approaches kinetic freeze-out.

\begin{figure}
\includegraphics[scale=0.4]{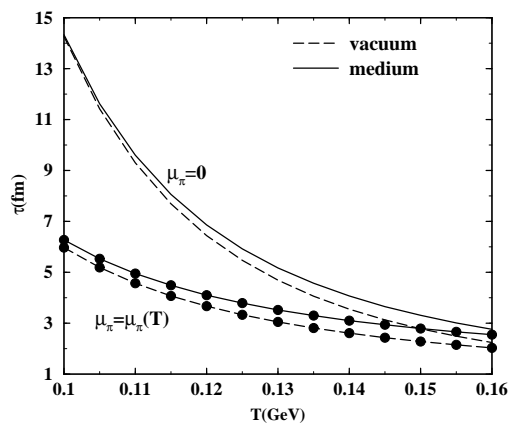}
\caption{The mean relaxation time with and without medium effects.}
\label{tau}
\end{figure}

At this stage it is worthwhile to compare the results with those obtained using
the so-called relaxation time approximation.
This method is the simplest way to linearize the
transport equation and is widely used. 
In this approach the 
distribution function $f(x,p)$ is assumed to go over to the equilibrium
distribution 
$f^{(0)}(x,p)$ over a time scale usually referred to as the relaxation 
time $\tau(p)$ which is actually given by the
inverse of the collision frequency $\omega(p)$. For 
a binary elastic collision $\pi(p)+\pi(k)\to\pi(p')+\pi(k')$ it is given by
\ba
\om(p)=&&\int d\Gamma_k\frac{\sqrt{s(s-4m_\pi^2)}}{E_p}f^{(0)}(E_k)(1+f^{(0)}(E_{p'}))\nonumber\\
&&\times(1+f^{(0)}(E_{k'}))\sigma(s)~.
\label{om_relax}
\ea

We plot in fig.~\ref{tau} the mean (thermal averaged) relaxation time as a 
function of temperature.
This is given by $\tau(T,\mu_\pi)=1/\overline\om(T,\mu_\pi)$ where
\be
\overline\om(T,\mu_\pi) = \int d^3p  f^{(0)}(p) \om(p)/\int d^3p  f^{(0)}(p)~.                                                                     
\ee                                                                                
The lower set of curves with filled circles correspond to a temperature dependent
chemical potential. The large difference with the upper set of curves depicting
the situation at vanishing pion chemical potential especially at lower temperatures shows
the role played by $\mu_\pi$. Accounting for the isospin degeneracy, the vacuum result for $\mu_\pi=0$ agrees with the
estimate of~\cite{Prakash,Davesne}. The solid line in both cases show a
noticeable medium effect compared to the vacuum. 

It may be pointed out that the mean relaxation time characterizes the rate of
change of the distribution function due to collisions and only serves as a
orientational guide to equilibrium~\cite{Prakash}. 
On the other hand the relaxation time of flows give the time scales
over which momenta and heat are transported. They cannot be obtained in the
Chapman-Enskog formalism where the neglect of all gradients of flows in the
conservation laws lead to infinite speeds for the flows~\cite{Davesne}.

\begin{figure}
\includegraphics[scale=0.4]{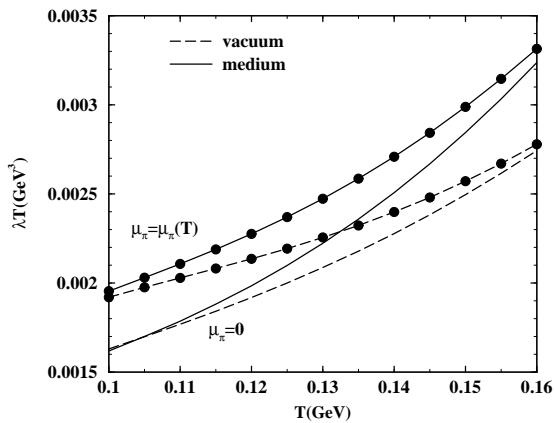}
\caption{$\lambda T$ as a function of $T$. in the relaxation-time approximation. 
The set of curves with filled circles correspond
to calculations done using a temperature dependent pion chemical potential.}
\label{relax}
\end{figure}

The transport equation in the relaxation time approximation reduces to
\be
\frac{\partial f}{\partial t}+\vec v_p\cdot\vec\nabla f=-\frac{(f-f^{(0)})}{\tau}.
\label{relax1}
\ee 
from which the thermal conductivity comes out to
be~\cite{Gavin},
\be
\lambda=\frac{2}{3T^2} \int d\Gamma_p\frac{p^2}{E_p}(E_p-h)^2\tau(p)
f^{(0)}(E_p)(1+f^{(0)}(E_p)).
\label{lambda-relax}
\ee

In fig.~\ref{relax} we have plotted $\lambda T$ versus $T$ both for zero and a 
temperature dependent chemical potential.
The substantial effect of the
medium is distinctly visible through the difference between the dashed and solid
lines in the two sets. The separation between the set of curves 
with and without circles shows the effect of the pion chemical potential
and as expected, is more at lower temperatures. 

The value of $\lambda$ for the various cases displayed in
figs.~\ref{muT} and \ref{relax} lie within $\sim$0.4-1.2 in units of fm$^{-2}$ at
$T=160$ MeV. Taking the peak value of the $\pi\pi$ cross-section  as shown
in fig.~\ref{sigmafig} these values are within reasonable agreement with those
of~\cite{Greif}.

To summarize, we have evaluated the thermal conductivity of an interacting pion
gas by solving the relativistic transport equation in the Chapman-Enskog and
relaxation time approximations. In-medium effects on the $\pi\pi$ cross-section
are incorporated through one-loop self-energies of the exchanged $\rho$ and
$\sigma$ mesons calculated using thermal field theory. The effect of chemical
freezeout is incorporated through a temperature dependent pion chemical 
potential which keeps the pion number conserved. It is observed that the
temperature dependence of the thermal conductivity is significantly affected.
It will be interesting to observe the consequences on the evolution of the late
stages of heavy ion collisions by including it in fluid-dynamical simulations. 

It may be pointed out that a realistic hadron gas is composed of
several types of hadrons and in principle should be considered 
for the evaluation of transport coefficients. However, treating the $\pi N$ gas
as a binary hadronic mixture the viscosities 
and thermal conductivities were found~\cite{Prakash} to be close to
those of a pion gas
due to the small concentration of nucleons.  
It may be worthwhile to investigate the role of medium effects in such systems
especially for situations involving high baryon density.

\end{document}